\documentclass[final,3p,times,singlecolumn]{elsarticle}
\usepackage{graphicx}
\usepackage{amsmath}
\usepackage{amssymb}
\usepackage{hyperref}

\begin{document}

\title{Physical Properties of Nano-crystalline Sm$_2$CoMnO$_6$: structure, magnetism, spin-phonon coupling and dielectric study}

\author{Ilyas Noor Bhatti\corref{cor2}}\address{Department of Physics, Jamia Millia Islamia University, New Delhi - 110025, India.}
\ead{inoorbhatti@gmail.com}
\author{Imtiaz Noor Bhatti}\address{Department of School Education, Government of Jammu and Kashmir, India.}

\author{Rabindra Nath Mahato}\address{School of Physical Sciences, Jawaharlal Nehru University, New Delhi - 110067, India.}
\author{M. A. H. Ahsan}\address{Department of Physics, Jamia Millia Islamia University, New Delhi - 110025, India.}

\begin{abstract}
  Structural, magnetic and dielectric properties of Sm$_2$CoMnO$_6$ have been studied. X-ray diffraction and Rietveld analysis show that the sample crystallizes in the monoclinic structure with \textit{P2$_1$/n} space group. Magnetic study shows that the sample undergoes a paramagnetic to ferromagnetic phase transition around $T_c$ $\sim$148 K and a low temperature ordering of Sm$^{3+}$ moments. Temperature dependent Raman study shows spin-phonon coupling present in this material. Dielectric response of Sm$_2$CoMnO$_6$ shows strong frequency dependence with large dispersion and large dielectric constant. Deviated from ideal Debye's model is found in impedance spectroscopy. Further, the activation energy obtained from tangent loss and complex impedance suggests that the relaxation process is due to electron hopping. AC conductivity has been studied to understand the conduction mechanism. The detailed analysis of AC conductivity shows that quantum mechanical tunneling of charges is involved in the conduction mechanism. The exponent obtained from AC conductivity measurements confirms the non-Debye's nature of Sm$_2$CoMnO$_6$.
\end{abstract}

\maketitle
\section{Introduction}
Materials with coupled magnetic and dielectric properties have become the topic of intensive research in recent times. Double perovskite material have shown potential as promising candidate in this research field and have exhibited novel phenomenon such as ferromagnetism, magneto-dielectric, multiferroic, magnetoelectric, ferromagnetic semiconductors etc.\cite{ndi, vilar, yang, dass, ndia, de, ghara, ilyas4} However, the magnetic and electric coupling is rather weak and appears at low temperature thus making it difficult to use for application purpose. The ferromagnetic ordering along with strong insulating behavior of some 3$d$ based double perovskite is eye caching and can exhibit exotic phenomenon.\cite{fuh, azu}

Recently, several double perovskites have been extensively studied and have exhibited exotic phenomena. For instance, partially disordered ferromagnetic semiconductor La$_2$NiMnO$_6$ reportedly exhibited colossal magnetodielectricity and multiglass properties.\cite{choudhary} This material also exhibits magnetocapacitance close to room temperature\cite{rogado} and exhibit novel behavior with doping.\cite{hoss2} Lu$_2$CoMnO$_6$ with up-up-down-down ($\uparrow\uparrow\downarrow\downarrow$) magnetic structure exhibits ferroelectricity which originates from the exchange-striction mechanism.\cite{zhang} La$_2$Mn(Ni/Fe)O$_6$ shows ferromagnetic ordering and high refrigerant capacity.\cite{gau} In these compounds the Ni/Mn are ferromagnetically ordered via super exchange interactions mediated by oxygen. Materials with Co/Mn at B$^\prime$/B$^\prime\prime$ are again an interesting class of double perovskite materials with interesting physical properties. Double provskit material like La$_2$CoMnO$_6$ shwows ferromagnetic behavior around $\sim$220 K and shows magnetocapacitance and these properties draw the interest of researches to investigate Co/Mn based perovskites.  Tb$_2$CoMnO$_6$ is an anisotropic magnetic material and shows giant rotating magnetocaloric effect \cite{moon}. Further, pyroelectric/magnetoelectric properties along with ferromagnetic transition have been observed for (Y/Er)$_2$CoMnO$_6$.\cite{bisco1, bisco2} Structural and magnetic properties have been studied in single crystal (Ho,Tm,Yb,Lu)$_2$CoMnO$_6$ materials using neutron diffraction and DC magnetization.\cite{bisco3} Mean field type ferromagnetism and occurrence of Griffith phase has also been observed in nano-crystaline Pr$_2$CoMnO$_6$.\cite{ilyas1, ilyas2} Raman study on bulk (Pr, Ho)$_2$CoMnO$_6$ reveals strong spin-phonon coupling in these similar materials.\cite{wliu, ilyas3, silva} Further, in bulk double perovskites dielectric properties have been reported and found thermally activated relaxation mechanism and non-debye's behavor.\cite{nath,jwchen, ilyas3, jgjwang} Despite so many interesting properties of Co/Mn based double perovskites these materials have not been studied in much detail especially in thin films and nano-crystalline form. 

Here we have chosen Sm$_2$CoMnO$_6$ to investigate its physical properties in nano crystalline form. It is worth noting that the charge state Co$^{2+}$ and Mn$^{4+}$  give an ordered monoclinic phase structure whereas $Co^{3+}$ and $Mn^{3+}$ would result in B-site disordered phase. Further, in ordered phase of Sm$_2$CoMnO$_6$ the Co$^{2+}$-O-Mn$^{4+}$ would couple via superexchange interaction and give rise to a ferromagnetic ordering.\cite{good} Whereas for disordered phase moments favour antiferromagnetic spin arrangment. Furthermore, it is expected that with decreasing ionic radii of rare earth ions at A-site the structrual modification will take place and magnetic ordering temperature decreases. Sm$^{3+}$ ion is smaller than the La$^{3+}$  and Pr$^{3+}$ ions we expect reduced magnetic phase transition temperature.\cite{vasi} Further we investigate nano-crystalline Sm$_2$CoMnO$_6$ sample so the effect of grain size will also effect the physical properties specially dielectric behavior of sample. 

In this paper, we have studied nano-crystalline sol-gel prepared Sm$_2$CoMnO$_6$. Structural, magnetic, dielectric and transport properties have been studied in detail. Our aim in to investigate the physical properties of nano-crystalline Sm$_2$CoMnO$_6$. Further, we will compare the obtained result to results obtained for bulk material in earlier literature. Magnetic ordering, spin-phonon coupling and dielectric response are much focused in this study.

\begin{figure}[th]
	\centering
		\includegraphics[width=8cm, height=12cm]{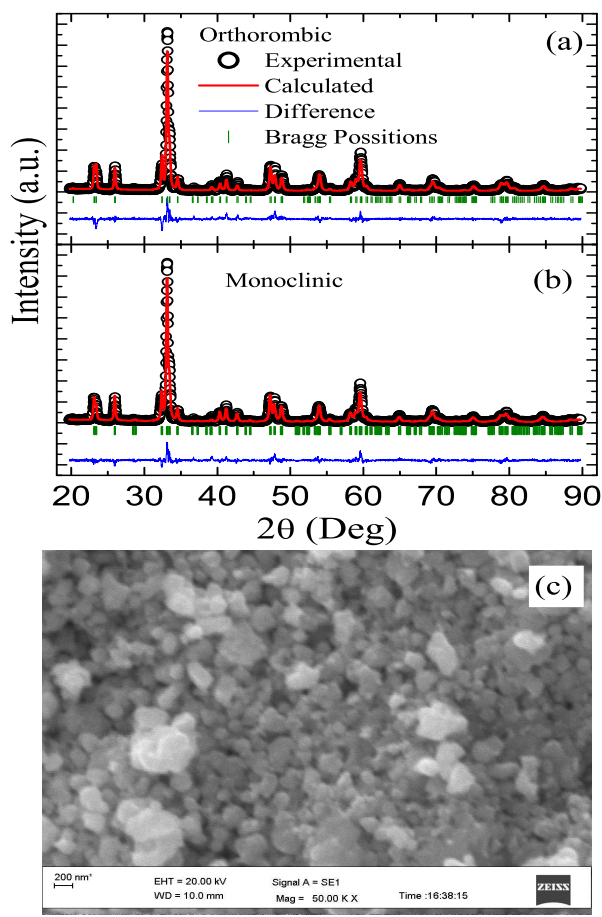}
\caption{(Color online) Comparative x-ray diffraction analysis by Rietveld refinement for nano-crystalline Sm$_2$CoMnO$_6$ (a) orthorhombic, (b) monoclinic phase. (c) SEM micrograph for nano-crystalline Sm$_2$CoMnO$_6$.}
	\label{fig:Fig1}
\end{figure}

\section{Experimental details}
The nano-crystalline sample of Sm$_2$CoMnO$_6$ was prepared by the sol-gel method as adopted in earlier work.\cite{ilyas1}  Sm$_2$O$_3$, Co(NO$_3$)$_2$.6H$_2$O and C$_4$H$_6$MnO$_4$.4H$_2$O  were taken in the stoichiometric ratio and dissolved in water. Sm$_2$O$_3$ is insoluble in the water, so we added HNO$_3$ drop by drop with continuous stirring on a magnetic stirrer with heating 70 $^o$C temperature until a clear solution is obtained. These solutions were then poured into a 400 ml beaker half-filled with the citric acid solution in water. The beaker with the obtained solution was then placed on the magnetic satire at 90 $^o$C temperature for 24 hours until the gel is formed. Then the temperature was increased to 200 $^o$C to dry up the gel. The sample was then collected from the beaker and crushed into fine powder by grinding in mortar and pestle. For the phase purity of the sample, the fine powder was then heated in a furnace at 800 $^o$C and 900 $^o$C for 12 hours.

The x-ray diffraction was taken for structural characterization and to confirm the phase purity of the sample using Rigaku made differectometer MiniFlex600 . XRD data was collected at room temperature in 2$\theta$ range 10$^o$ -90$^o$ with step size of 0.02$^o$ and scan rate of 2$^o$/min. Crystal structural analysis was done by  Rietveld refinement of XRD data using Fullprof program. The XPS measurements were performed with base pressure in the range of $10^{-10}$ mbar using a commercial electron energy analyzer (Omnicron nanotechnology) and a non-monochromatic Al$_{K\alpha}$ X-ray source (h$\nu$ = 1486.6 eV). The XPSpeakfit software was used to analyze the XPS data. The samples used for XPS study are in pallet form where an ion beam sputtering has been done on the samples to expose clean surface before measurements. Magnetic data was collected by Physical Properties Measurement System by Cryogenic Inc. Labram-HR800 micro-Raman spectrometer with diode laser having wavelength ($\lambda$) = 473 nm have been used to record the temperature dependent Raman spectra. This spectrometer use grating with 1800 groves/mm and CCD detector with a high resolution of $\sim$ 1 cm$^{-1}$. A THMS600 stage from Linkam UK have been used for temperature variation with the stability of $\pm$0.1 K for the low temperature Raman measurements. Dielectric measurements in the frequency range from 1 Hz to 5.6 MHz were performed using a computer controlled dielectric spectrometer.

\section{Result and Discussions}

\subsection{Structural study}
X-ray study was done for structural characterization of Sm$_2$CoMnO$_6$ and data was recorded.  Further detailed structural study of the sample was done by Rietveld refinement of XRD data using Fullprof program.  XRD patterns along with Rietveld refinement for orthorhombic and monoclinic phases are shown in the Fig. 1a and 1b. We have performed Rietveld refinement of XRD data for both orthorhombic and monoclinic crystal structures however the fitting for orthorhombic phase is not as good as for monoclinic. The R-factor R$_{exp}$, R$_{wp}$  and $\chi^{2}$ for orthorhombic structure comes out 15.0, 19.7 and  1.78 respectively. However, the rietveld refinement of XRD pattern with monoclinic phase gives reasonable good fitting  with R$_{wp}$, R$_{exp}$ and $\chi^{2}$  values as 18.6, 14 and 1.5 respectively. These values are acceptable and shows sample is chemically pure and single phase.\cite{bhatti1, bhatti2} Further, results confirm Sm$_2$CoMnO$_6$ sample adopt monoclinic structure with \textit{P2$_1$/n} space group. The sample has unit cell parameters are $a$ = 5.330624 $\AA$, $b$ = 5.498045 $\AA$, $c$ =  7.566482 $\AA$ and $\beta$ is  89.98(5) $^o$ with unit cell volume is 221.75 $\AA^3$. Further, to obtain the particle size in the nano-crystalline sample of Sm$_2$CoMnO$_6$ we have performed the scanning electron microscopy. Fig. 1c shows the SEM image obtained for the nano-crystalline structure. The SEM image is analyzed using ImageJ software. We observed that the average crystallite size for present compound is $\sim$84.17 nm.

\begin{figure}[t]
	\centering
		\includegraphics[width=8cm, height=12cm]{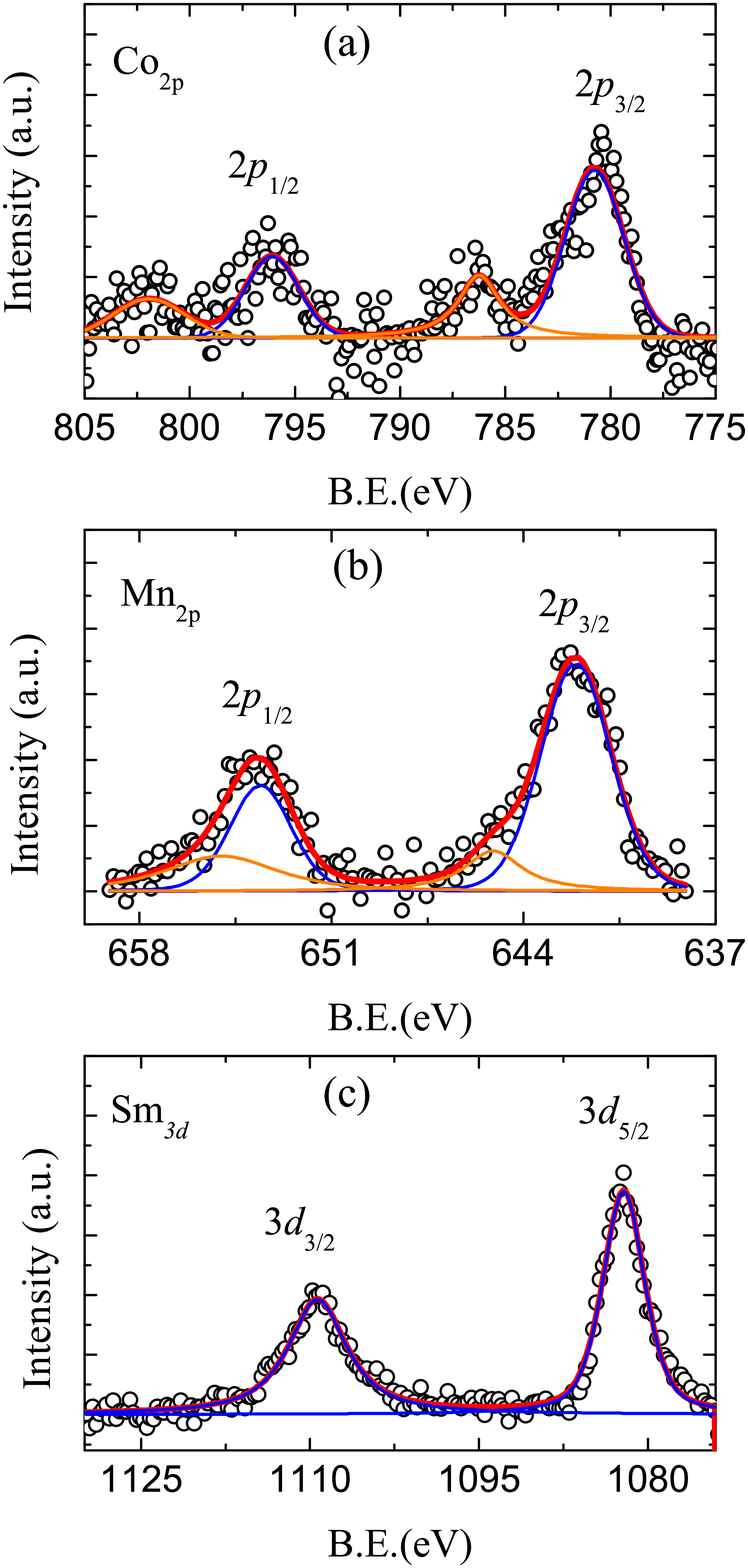}
\caption{(Color online) (a) The XPS core level spectra of Co 2$p$. (b) XPS core level spectra of Mn 2$p$. (c) XPS core level spectra of Sm 3$d$. In the figure the red solid line is the overall envolp of the XPS spectrum and the other colored solid lines are the other respective fitted peaks.}
	\label{fig:Fig2}
\end{figure}

\subsection{X-ray photo-electron spectroscopy (XPS)}
The physical properties of a compound is largely described by the oxidation state of cations present in material. XPS is a vital tool to understand the charge state of cations. We have employed the XPS to study the cationic charge state of Co, Mn and Sm in Sm$_2$CoMnO$_6$. XPS spectrum of Co 2$p$ is shown in Fig. 2a, it is evident that there are two peaks located at 780.7 eV and 796.01 eV for CO 2$p$$_{3/2}$ and Co 2$p$$_{1/2}$  respectively resulting from spin orbital splitting of 2$p$ orbital with 15.3 eV. Beside the Co 2$p$ peaks two satellite peaks have also been observed close to Co 2$p$ peaks and are in agreement resulting with in literature.\cite{wang, qiu, xia}  The peaks locations of Co 2$p$ core level indicates that the Co cations are presents in +2 oxidation states.

The measured XPS spectrum for Mn 2$p$ core levels along with peak fitting is shown in Fig. 2b. The Mn 2$p$ spectrum shows two distinct peaks located at 642 eV and 654 eV corresponding to Mn 2$p$$_{3/2}$ and 2$p$$_{1/2}$ resulted from spin orbital splitting of 2$p$ orbitals with splitting energy of 12 eV. There are two small peaks at the higher energy side is observed in fit the data and believed due to satellite correction. The resulting corroborates with the reported literature.\cite{ida, sachoo, cao} The peak position reveals that the Mn cation is present in +4 oxidation state. 

XPS spectra of Sm 3$d$ core level along with the fitting of peaks are shown in Fig. 2c. It is quite evident from the figure that the two distinct spin-orbital split peaks Sm 3$d$$_{5/2}$ and Sm 3$d$$_{3/2}$ are located at 1082.16 eV and 1109.33 eV respectively with at spin-orbital splitting energy of 27.17 eV. The detail analysis of XPS spectrum reveals that the Sm cation are present in +3 oxidation states as reported in literature.\cite{qliu, duan} Thus the XPS study reveals the oxidation state and the spin orbital splitting energy of the compounds present in the material. The observed oxidation state Co$^{2+}$, Mn$^{4+}$ and Sm$^{3+}$ suggests the crystallization of ordered phase of Sm$_2$CoMnO$_6$. Further, in the XPS data we do not see any contribution from other cationic valencies corresponding to the peak positions for Mn$^{3+}$ or Co$^{3+}$ unlike seen in the literature.\cite{qiu, xia, sachoo, cao, haung} 

\begin{figure}[th]
	\centering
		\includegraphics[width=8cm]{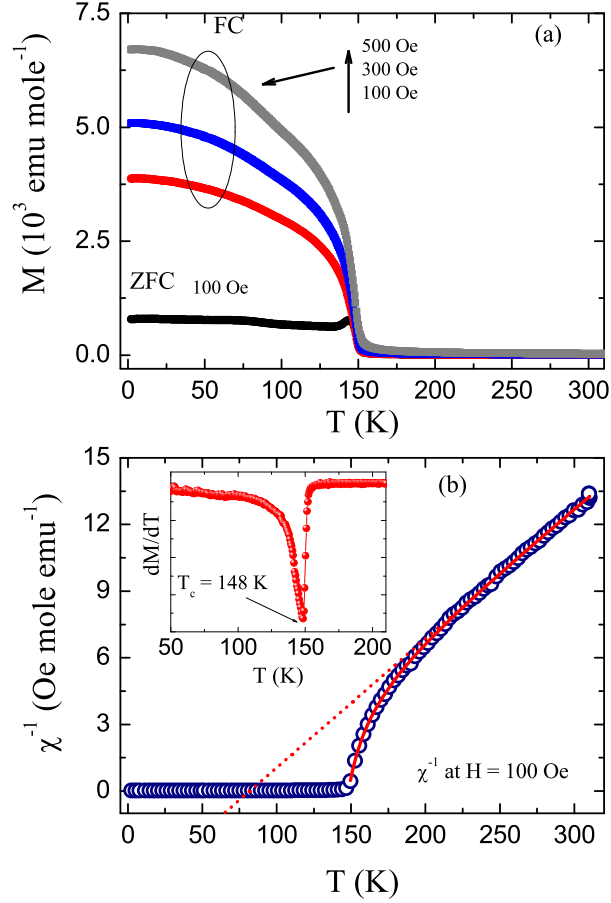}
\caption{(Color online) (a) Temperature dependent magnetization data $M(T)$ shown for Sm$_2$CoMnO$_6$ measured at different fields. (b) $M(T)$ data plotted in terms of inverse susceptibility $\chi^{-1}$, solid line is fitting due to modified Curie Weiss Law and dotted line shows conventional Curie Weiss Law fit. Inset shows dM/dT vs $T$ plot showing $T_c$ for Sm$_2$CoMnO$_6$.}
	\label{fig:Fig3}
\end{figure}

\begin{figure}[t]
	\centering
		\includegraphics[width=8cm]{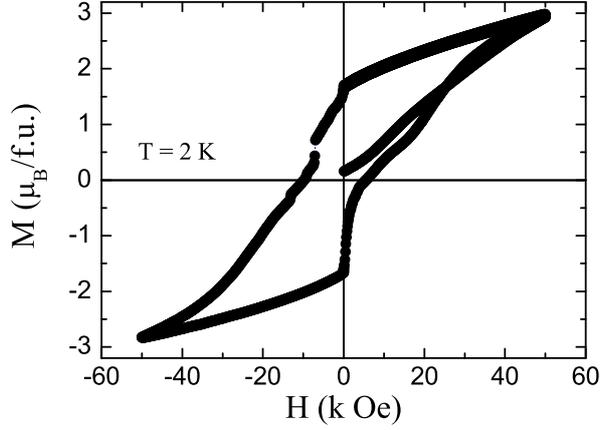}
\caption{(Color online) Isothermal magnetization $M(H)$ data collected at 2 K in applied field of $\pm$ 50 kOe for Sm$_2$CoMnO$_6$.}
	\label{fig:Fig4}
\end{figure}

\subsection{Magnetization study}
Fig. 3a shows the temperature dependent magnetization ($M(T)$) data measured in both zero field cooled ($ZFC$) and field cooled ($FC$) mode. $ZFC$ is measured in 300 K to 2 K range in an applied field of 100 Oe whereas $FC$ data is measure at three different applied magnetic fields (see Fig. 3a). It is quite evident that with decreasing temperature the magnetic moment ($M$) in $M(T)$ is steady till 160 K, however, with further decrease in temperature the magnetic moment begin to rise sharply. It is well known that the ordered Sm$_2$CoMnO$_6$ double perovskite is expected to exhibit ferromagnetic ordering. Since, from XPS study we observed that Mn and Co cations are in +4 and +2 oxidation states respectively with outer electronic configuration Mn$^{4+}$ (t$_{2g}^{3}$e$_{g}^{0}$), Co$^{2+}$ (t$_{2g}^{5}$e$_{g}^{2}$) these cations take part in superexchange interaction and are ferromagnetically ordered.  The sharp rise in moment below 160 K is marked by paramagnetic to ferromagnetic phase transition in Sm$_2$CoMnO$_6$. With further decrease in temperature, large bifurcation in $M_{ZFC}$ and $M_{FC}$ appears. $M_{ZFC}$ curve shows a peak like behavior around $T_c$ as shown in Fig. 3a, with further decreasing temperature the moment remain steady. However, $M_{FC}$ shows a typical ferromagnetic like feature. The nano-crystalline Sm$_2$CoMnO$_6$ shows a magnetic phase transition from PM-FM with T$_c$ $\sim$148 K, as obtained from point of inflection in $dM/dT$ vs $T$ plot shown in inset Fig. 3b. We have measured $M_{FC}$ at different applied field and observe that with increasing field, $T_c$ shifts to higher field which is due to forced spin arrangement in field direction. 

To understand the magnetic behavior in further detailed we have plotted magnetization data in terms of temperature dependent inverse magnetic susceptibility i.e. $\chi^{^{-1}}$ vs $T$ as shown in Fig. 3b. In paramagnetic region close to above $T_c$ we observe that the inverse magnetic susceptibility ($\chi^{^{-1}}$) shows deviation from linearity unlike as expected for paramagnetic system by Curie Weiss law (see dotted line). However, such deviation in $\chi^{^{-1}}$ close to $T_c$  is observed and expected for double perovskite compounds which can be understood by using modified Curie Weiss law described as:\cite{booth}

\begin{eqnarray}
\chi = \frac{C_{TM}}{T - \theta_{TM}} + \frac{C_{RE}}{T - \theta_{RE}}
\end{eqnarray}

where $C_{TM}$, $C_{RE}$ and $\theta_{TM}$,  $\theta_{RE}$ are the Curie Constants and paramagnetic Curie temperatures respectively, here subscript TM and RE represent transition metal ions and rare earth ions. We have fitted the susceptibility data of  Sm$_2$CoMnO$_6$ at high temperature above $T_c$. The fitting parameters obtained from fitting of $\chi^{-1}$ with Eq. 1 in Fig. 3b were used to calculate Curie constant and $\theta_P$. The values obtained for $C_{TM}$ and  $\theta_{TM}$ for Sm$_2$CoMnO$_6$ are 3.081 emu K$^{-1}$ mole$^{-1}$ Oe$^{-1}$ and 146.96 K respectively. The large positive value of $\theta_P$ signifies that the ferromagnetic ordering is present in the sample. Further, effective magnetic moment is calculated using formula  $\mu_{eff}$ = $\sqrt{3 C_{TM} k_{B}/N}$ where $C_{TM}$ is obtained from modified Curie Weiss fitting in Fig. 3b. The value of $\mu_{eff}$ is 4.91474 $\mu_B$/f.u. Further, from fitting we have obtained the $C_{RE}$ value for rare earth ions as 15.953 emu K$^{-1}$ mole$^{-1}$ Oe$^{-1}$ which is close to the value of Sm$^{3+}$ free ions. 

Isothermal magnetization $M(H)$ data have been collected at 2 K up to $\pm$50 kOe applied magnetic field as shown in Fig. 4. $M(H)$ curve shows hysteresis which is signature of ferromagnetic ordering. However the $M(H)$ curve is not symmetric. Further the magnetic moment does not show any signature of saturation even at highest applied magnetic field of 50 kOe. The magnetic moment at 50 kOe is about 3.1 $\mu$$_B$/f.u. where as the remanent magnetization and coercive force is 1.7 $\mu$$_B$/f.u. and 10 kOe respectively. 

Reducing particle size in nano-scale does effect the magnetic properties of compounds in great deal. Many materials with particles dimensions at nano-scale shows drastic change in magnetic properties in comparison to their bulk form, especially manganites show superparamagnetism, spin-glass behavior, core-shall spin structure etc.\cite{tzhang} It is well established that the reducing particle size effect unit cell dimensions, which effect the Metal-Oxygen bond length and bond angles hence effect the magnetic properties.\cite{rer}  Finally, to understand the effect of nano-size on magnetism we made a comparison of the magnetic parameters with bulk as reported in literature.\cite{nath} It is found that in nano form Sm$_2$CoMnO$_6$ sample have large Curie temperature 146.96 K whereas for bulk it is found to be around 122 K \cite{nath}. The experimentally obtained value of effective magnetic moment found to be lower than bulk as reported.\cite{nath} The effect on magnetic properties can be described due to surface ferromagnetism on small nano particles as found in case of doped perovskite.\cite{hoss1}

\begin{figure*}[t]
	\centering
		\includegraphics[width=14cm]{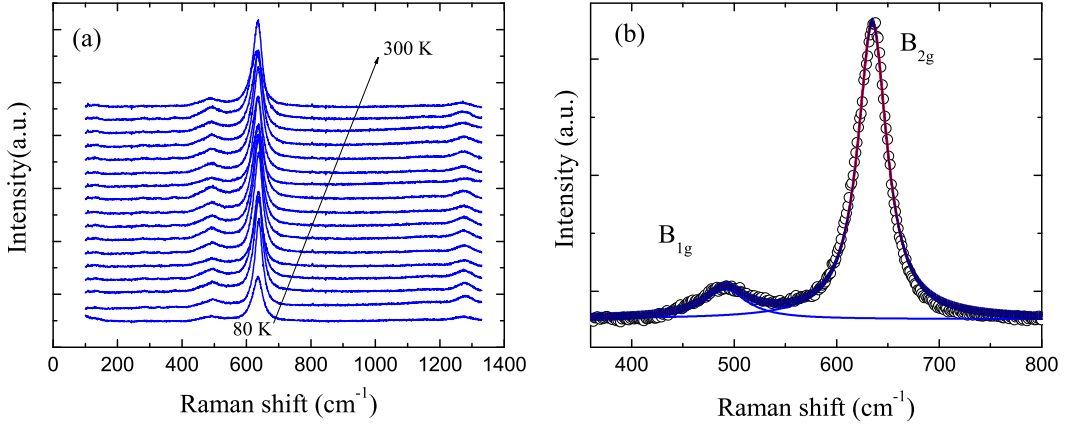}
\caption{(Color online) (a) Raman spectra of Sm$_2$CoMnO$_6$ measured at different temperatures.   (b) shows the line shape and its Lorentzian fitting of A$_{1g}$ and B$_{2g}$ Raman modes at 496 and 636 cm$^{-1}$ respectively.}
	\label{fig:Fig5}
\end{figure*}

\begin{figure}[t]
	\centering
		\includegraphics[width=8cm]{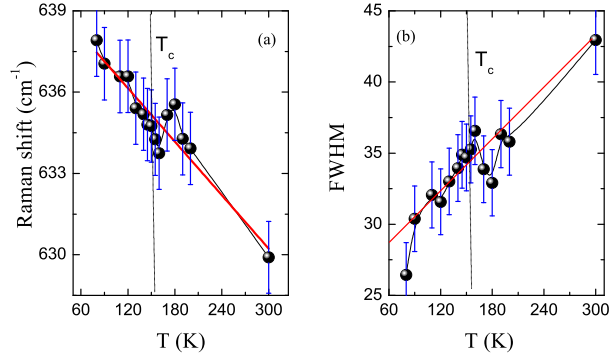}
\caption{(Color online) Temperature Variation of (a) Raman shift (b) FWHM for Raman mode at 636 cm$^{-1}$ corresponding to stretching of Co/MnO$_6$ for Sm$_2$CoMnO$_6$. The solid line is fitting due to Eq. 2}
	\label{fig:Fig6}
\end{figure}

\subsection{Temperature dependent Raman study}

Fig. 5a shows the temperature dependent Raman spectra taken at selective temperatures across magnetic transition. The Raman spectrum are taken at 10 K, room temperature and across the magnetic transition at close temperature intervals. It is evident from the Fig. 5a the important feature in the Raman spectra are the prominent Raman modes at 636 and 496 cm$^{-1}$ corresponding to B$_{2g}$ stretching mode and A$_{1g}$ breathing mode respectively. These Raman modes are due to stretching, bending and rotation of (Co/Mn)O$_6$ octhadera. It is known from the theoretical lattice dynamics that the strong sharp peak at 636 cm$^{-1}$ originates from a symmetric stretching of the (Co/Mn)O$_6$ octahedra, whereas the band at around 496 cm$^{-1}$ describes a mixed type vibration of antisymmetric stretching and bending.\cite{ilive} Additionally, the modes at $\sim$1278 cm$^{-1}$  represent the second-order overtones of the breathing mode.\cite{meyer} The temperature dependent Raman spectra shows modification in intensity of peaks well as their peak positions corresponding to each modes. Further, it is well known that the behavior of ordered and disordered phases in double perovskite can be understand using Raman spectra, such detail study have been carried out on La$_2$CoMnO$_6$ where mode at 636 cm$^{-1}$ seems prominent and assigned to a monoclinic P2$_1$/n structure.\cite{ilive, meyer} The present study shows similar frequency results. 

To understand the presence of spin phonon coupling in present compound, we have analyzed the Raman data of present series with the following anharmonic decay model:\cite{harish}

\begin{eqnarray}
\omega(T) = \omega{_0} - A\left[1 + \frac{2}{exp\left(\frac{\hbar\omega_0}{2k_BT}\right) - 1}\right]
\end{eqnarray}

\begin{eqnarray}
\Gamma(T) = \Gamma{_0} - B\left[1 + \frac{2}{exp\left(\frac{\hbar\omega_0}{2k_BT}\right) - 1}\right]
\end{eqnarray}

where $\omega_0$ and $\Gamma_0$ are the intrinsic frequency  and line width of the optical mode, A and B are the anharmonic coefficients. $\omega(T)$  and $\Gamma(T)$ describes expected temperature dependence of a phonon mode frequency and line width  due to anharmonic phonon-phonon scattering. 

The temperature dependent Raman spectra shown in Fig. 5a have been analyzed using Lorentz functions. From the fitting of these spectra, peak position of Raman modes and line width have been obtained. Fig. 6a shows the temperature dependent peak position of Raman mode corresponding to stretching mode. The mode frequency increases with decreasing temperature, and shows a deviation around 150 K which is magnetic phase transition and further increases down to lowest measured temperature. To understand further we have fitted the phonon mode frequency with Eq.2 anharmonic decay model (solid red line in Fig 6a). We observed that around $T_c$ phonon frequency shows a deviation from anharmonic behavior and such behavior is due to spin phonon coupling present in materials. Such results have been reported in many other materials where at magnetic transition the material shows strong deviation from anharmonic behavior.\cite{sandi, grana, lave} Similarly, the line width as a function of temperature is plotted and shown in Fig. 6b. The line width is also fitted with the anharmonic model given in Eq. 2. It is found that the line width shows a deviation across $T_c$ in a similar fashion as mode frequency does. The deviation of mode frequency and line width form anharmonic behavior around magnetic ordering temperature is due to the additional scattering resulting from spin-phonon interaction. 

\begin{figure*}[th]
	\centering
		\includegraphics[width=14cm]{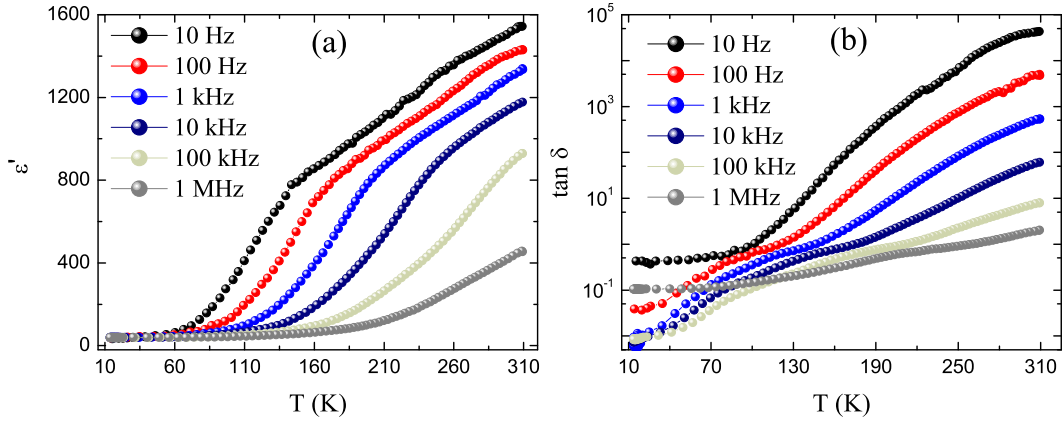}
\caption{(Color online) Temperature dependent (a) real part of complex dielectric permittivity ($\epsilon$$^\prime$) (b) loss tangent (tan $\delta$) measure for Sm$_2$CoMnO$_6$ in the temperature range of 20 K to 300 K at various frequencies.}
	\label{fig:Fig7}
\end{figure*}

\begin{figure}[th]
	\centering
		\includegraphics[width=8cm]{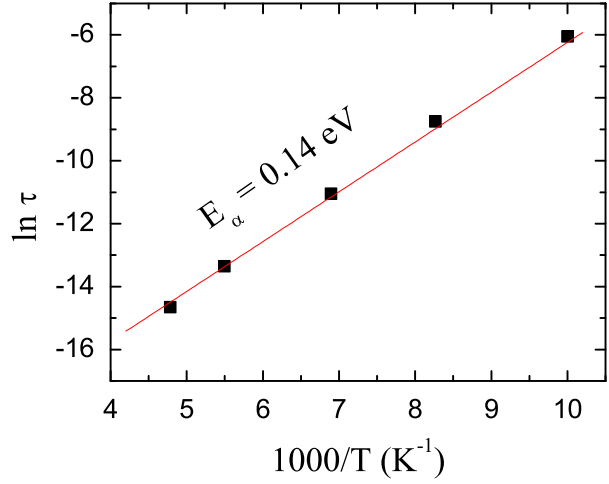}
\caption{(Color online) Variation of relaxation time against normalized temperature i.e ln $\tau$ vs 1000/T obtained from tengant loss in Fig. 7(b).}
	\label{fig:Fig8}
\end{figure}

\subsection{Dielectric study}

Fig. 7a and 7b represent the temperature dependent real and imaginary part of complex dielectric permittivity $\epsilon^{\prime}$ and tan $\delta$  respectively measured in the temperature range 20 K to 300 K at different frequencies for Sm$_2$CoMnO$_6$. Further, for relaxor systems with relaxation mechanisms, each relaxation component will correspond to plateaus in $\epsilon^{\prime}$(T) and respond with peaks in tan $\delta$. For this material we have observed that with increasing temperature the $\epsilon^{\prime}$ increases,  at low temperature the $\epsilon^{\prime}$ increases slowly however with increasing temperature $\epsilon^{\prime}$ increases sharply. The high response of dielectric permeability is due to grain boundary effect.\cite{jwchen, mansuri}  Further, with increasing frequencies $\epsilon^{\prime}$ decreases sharply. The higher value of $\epsilon^{\prime}$ at low frequency is attributed to the accumulation of the charges at grain boundaries  On careful observation of tangent loss curve tan $\delta$ it is clearly seen that there is a broad hump at low temperatures which is feature of relaxor phenomenon. The observed relaxation is frequency dependent and shift to higher temperature with increasing frequency. This clearly indicates that the relaxation peaks in the dielectric loss is due to thermally activated mechanisms. The resonance condition defined as $\omega_p$$\tau_p$ = 1 where $\omega$ = 2$\pi$$f$ is defined as resonance frequency.

The relaxation mechanisms and its origin can be analyzed by fitting the peaks in tan $\delta$ with the Arrhenius law given as follow:
\begin{eqnarray}
 \tau_{tan \delta} = \tau_0 exp(\frac{-E_{\alpha}}{K_B T})
\end{eqnarray}
\begin{eqnarray}
 \tau_{tan \delta} = \frac{1}{2 \pi f_{tan \delta}})
\end{eqnarray}

 where, $T$ is the temperature where peak occurs in tangent loss curve at a particular frequency $f_{tan \delta}$, $\tau_0$ and $E_{\alpha}$ are characteristic relaxation temperature and activation energy respectively and, $k_B$ is the Boltzmann constant. Fig. 8 shows the plot of dielectric loss peaks as a function of absolute temperature. From the fitting parameters of the data using Eq. 4 we have calculated activation energy $E_{\alpha}$ = 0.14 eV. 

\begin{figure*}[th]
	\centering
		\includegraphics[width=14cm]{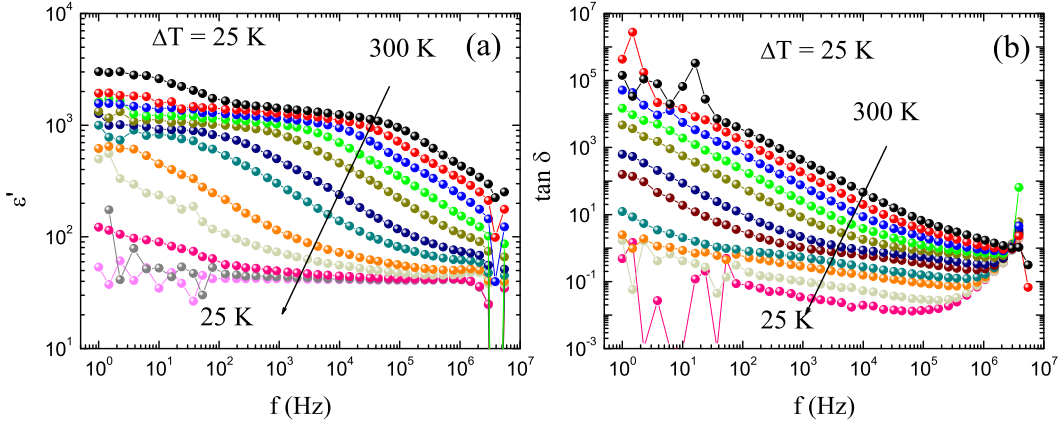}
\caption{(Color online) Frequency dependent (a) real part of complex dielectric permittivity ($\epsilon$$^\prime$) (b) loss tangent (tan $\delta$) measure for Sm$_2$CoMnO$_6$ at various temperatures between 25 K and 300 K in the frequency range 1 Hz to 5.6 MHz.}
	\label{fig:Fig9}
\end{figure*}

To further understand the dielectric response we have measured the frequency dependent $\epsilon$$^\prime$ and tan $\delta$ over the frequency range 1 Hz to 5.5 MHz for Sm$_2$CoMnO$_6$ at different temperatures. In the Fig. 9a we observed that Sm$_2$CoMnO$_6$ exhibits high dielectric constant at low frequency and at high temperature. The dielectric spectrum shown in Fig. 9a clearly shows two plateaus well separated by dispersion. The separate plateau in Fig. 9a are attributed to static and optical dielectric constant. Further, the frequency dependent dielectric constant shows the dispersion which moves to higher frequency with increasing temperature. In Fig. 9b we have shown the tan $\delta$ as a function of frequency at different temperature in the range of 20 K to 300 K. It is observed that at low frequency the loss is high which decreases with increasing frequency and decreasing  temperature.

\subsection{Impedance spectroscopy}

Impedance spectroscopy is vital and informative technique to understand and distinguish the contributions to the electric and dielectric properties from  grain, grain boundaries and electrode-sample contact interface. Various involve in relaxation mechanisms can be identified by plotting impedance in a complex plan at various temperature. The complex impedance describe by the equation:\cite{fang}

\begin{eqnarray}
Z^* = Z^\prime + jZ^{\prime\prime}
\end{eqnarray}

\begin{eqnarray}
 Z^\prime = \frac{R}{1 + (\omega\tau)^2}
\end{eqnarray}

\begin{eqnarray}
Z^{\prime\prime} = \frac{\omega R \tau}{1 + (\omega\tau)^2}
\end{eqnarray}
where Z$^*$ is complex impedance, Z$^\prime$ and Z$^\prime\prime$ are real and imaginary parts of impedance respectively. R is resistance, $\omega$ is angular frequency and $\tau$ is relaxaton time.
Fig. 10a shows the real part of complex impedance ($Z^{\prime}$) plotted as a function of frequency in the frequency range 1 Hz to 5.6 MHz at various temperatures between 50 K to 300 K. For clarity both the axis are in logarithmic scales. It is quite evident from the figure that $Z^{\prime}$ decreases with increasing temperature. At low temperature $Z^{\prime}$ gradually decreases with increasing frequency, however at temperatures above 100 K, $Z^{\prime}$ initially remains independent of frequency then at higher frequency it began to decreases. Further, the frequency independent region moves to higher frequency with increasing temperature. Further, it is observed that at higher frequency and high temperature $Z^{\prime}$ is almost similar. This feature is possibly due to the release of accumulated space charges at high temperatures hence contribute to the enhancement of conduction in this material at high temperature. Imaginary part of impedance ($Z^{\prime\prime}$) is shown in Fig. 10b for wide frequency range. $Z^{\prime\prime}$ shows a peak which attain $Z^{\prime\prime}_{max}$ for all the curves measured at different temperatures, further it is evident that the peak moves towards higher frequency with increasing temperature. The peak shift to higher frequency with increasing temperature suggests that the relaxation time constant decreases with increasing temperature. 

\begin{figure*}[th]
	\centering
		\includegraphics[width=14cm]{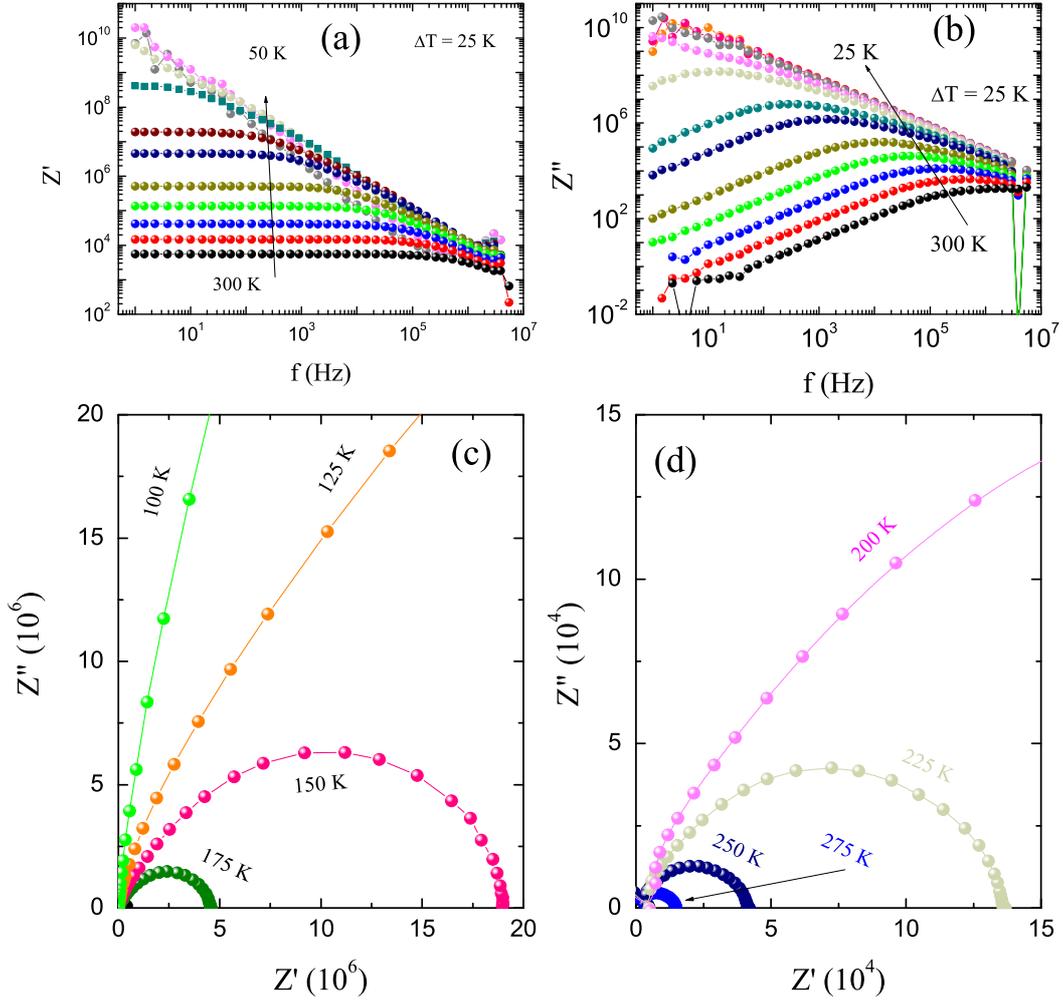}
\caption{(Color online) (a) Frequency dependent real part of impedance $Z^{\prime}$ measure at different temperatures. (b) Frequency dependent imaginary part of impedance $Z^{\prime\prime}$ measure at various temperatures. (c), (d) real $Z^{\prime}$ and imaginary part $Z^{\prime\prime}$ plotted in terms of Nyquist plot $Z^{\prime}$ vs $Z^{\prime\prime}$.}
	\label{fig:Fig10}
\end{figure*}

\begin{figure}[th]
	\centering
		\includegraphics[width=8cm]{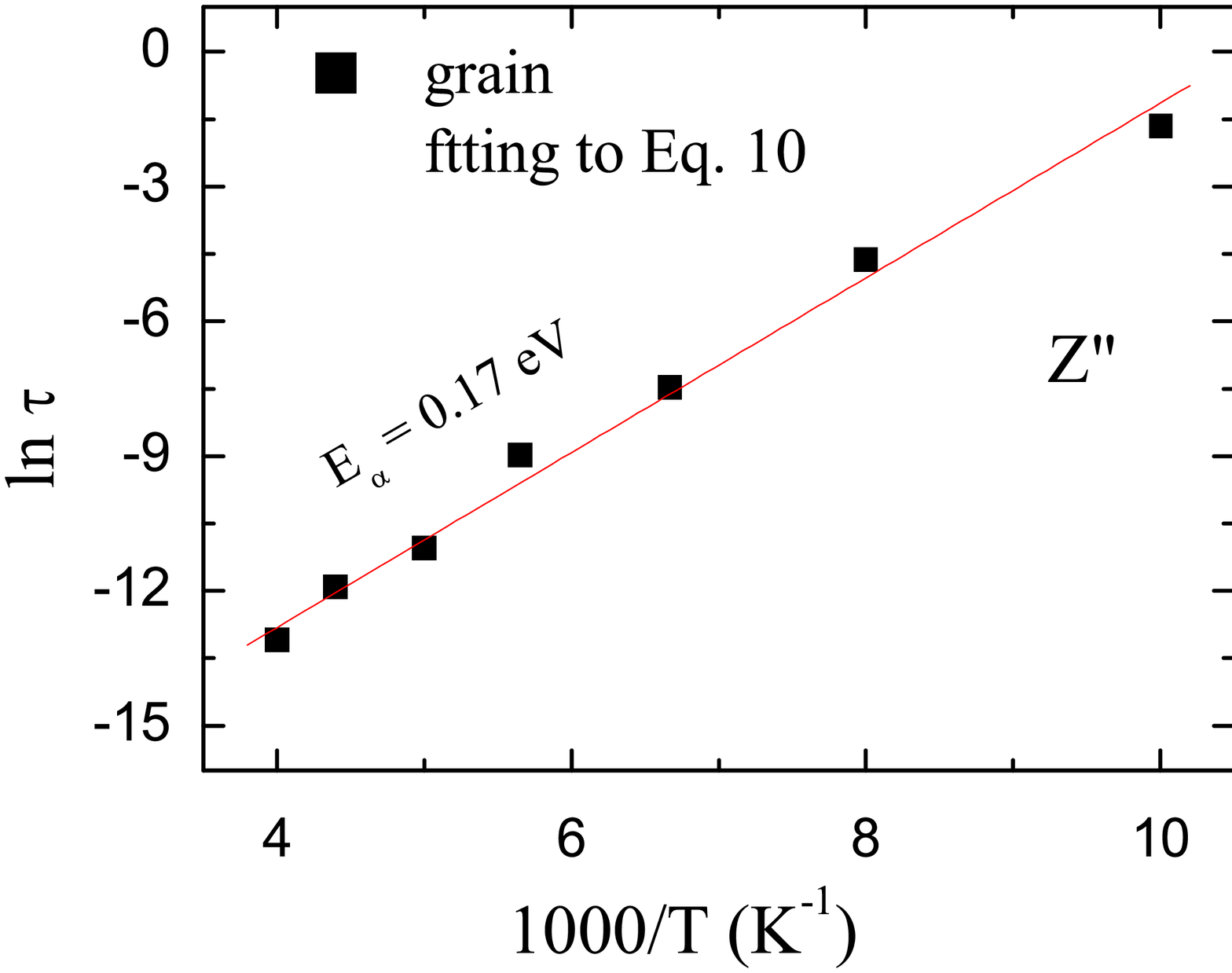}
\caption{(Color online) Variation of relaxation time against normalized temperature i.e ln $\tau$ vs 1000/T obtained from Z$^\prime\prime$ plot.}
	\label{fig:Fig11}
\end{figure}

We know that the most probable relaxation time ($\tau$) can be determined for relaxation system by identification of the position of the loss peak in the $Z^{\prime\prime}$ vs log ($f$) plots using the relation:

\begin{eqnarray}
\tau = \frac{1}{\omega} = \frac{1}{2 \pi f}
\end{eqnarray}

where $\tau$ is relaxation time and $f$ is the relaxation frequency. 
To further understand the relaxation behavior we have plotted the relaxation time $\tau$ vs inverse temperature 10$^3$/$T$ (K$^{-1}$). Fig. 11 shows the temperature variation of $\tau$, it is observed that the relaxation time follows Arrhenius behavior given as:
 
\begin{eqnarray}
\tau_b = \tau_0 exp \left( \frac{-E_\alpha}{k_BT} \right)
\end{eqnarray}

where $\tau_0$ is the pre-exponential factor, k$_B$ the Boltzmann constant and $T$ the absolutes temperature. The peak frequency $f_p$ obtained from the peak values of Z$^{\prime\prime}$ in Fig. 10b is plotted as relaxation time as a function of absolutes temperature shown in Fig. 11. The data is very well fitted with the Eq. 10 Arrhenius law for thermally activated relaxation mechanism. From the fitting parameters the activation energy (E$_{\alpha}$) have been calculated and is found to be 0.17 eV.

Fig. 10c and 10d shows the Z$^{\prime}$ vs Z$^{\prime\prime}$ in the form of Nyquist plots at some selective temperatures measure in the wide frequency range 1 Hz to 5.6 MHz. It is quite evident from the figures that the plot gives the semicircle in whole range of temperature. But the circle are compressed and their center are not on Z$^\prime$ axis. Which shows that the system is deviates from deal Debye's model, because for ideal Debye system the semicircles are supposed to have their Z$^\prime$ axis. The Nyqust plot give lot of information about grains, grain boundaries and electrode effect etc.  Further, we observed the semicircle in Nyquist plot shows a decrease in radius with increasing temperature which reflect that the resistivity decreases wth increasing temperature. Further the depression in semicircle are due to polarization effect. The non-Debye behavior account for grain boundaries, grain, stress and strain present in material. In this case we observed only one depress semicircle which indicate that the grain effect deviate the system from deal Debye behavior.

\begin{figure*}[th]
	\centering
		\includegraphics[width=14cm]{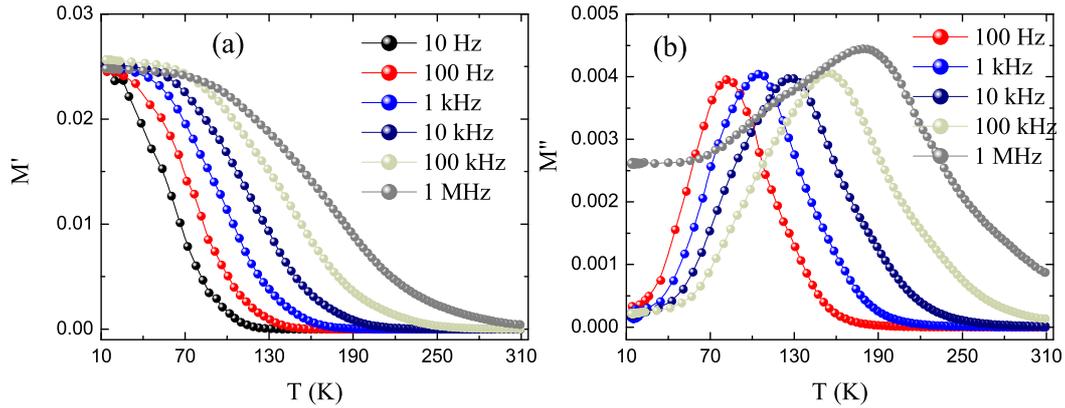}
\caption{(Color online) (a) Variation of real part of electrical modulus ($M^\prime$) with temperature. (b) Imaginary part of electrical modulus $M^{\prime\prime}$ as a function of temperature.}
	\label{fig:Fig12}
\end{figure*}

\begin{figure}[th]
	\centering
		\includegraphics[width=8cm]{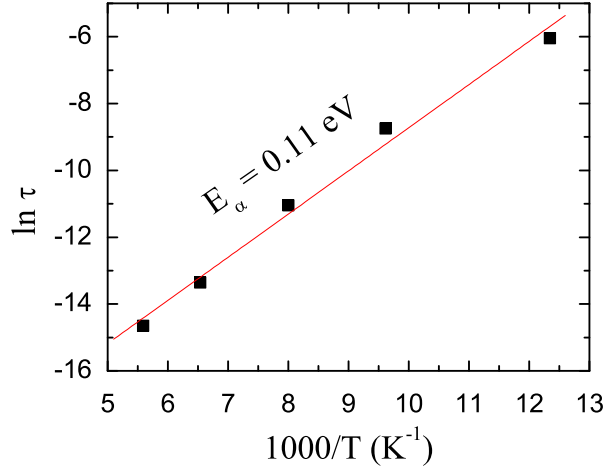}
\caption{(Color online) Variation of relaxation time against normalized temperature i.e ln $\tau$ vs 1000/T obtained from M$^\prime\prime$ plot.}
	\label{fig:Fig13}
\end{figure}

\subsection{Electric modulus}
Information of interface polarization, relaxation time, electrical conductivity and grain boundary
conduction effects etc can be deduced from electrical modulus of materials. Figs. 12a and 12b show the temperature dependent real (M$^{\prime}$) and imaginary (M$^{\prime\prime}$) part of electrical modulus obtained at selective frequencies for Sm$_2$CoMnO$_6$ in the temperature range of 20 K to 300 K.

Fig. 12a shows variation of real part of electrical modulus as a function of temperature at different frequencies. It is observed that M$^\prime$ ncreases wth decreasing temperature at low temperatures all the curves of M$^\prime$ measured at different frequencies merge in to one curve. Similar feature is observed at high temperatures also. There is large dispersion in M$^\prime$ at intermediate temperatures around 150 K which may be activated by magnetic ordering. The M$^\prime$ increases with increasing frequency, but with temperature the behavior remain same. Fig. 12b shows variation of $M^{\prime\prime}$ with frequency at selected temperatures. Once again, $M^{\prime\prime}$ spectroscopy plot reveals relaxation phenomena in the material. The maximum value ($M^{\prime\prime}$) of the $M^{\prime\prime}$ peak shifts to higher frequency, which suggests that hopping of charge carriers is predominantly thermally activated. Asymmetric broadening of the peak indicates spread of relaxation with different time constants, which once again suggests the material is non-Debye-type. In Fig. 13 we have plotted the $\tau$ vs absolute temperature $T$  calculated from the frequency for which peak value M$^{\prime\prime}_{max}$ appears in Fig. 12b by using the relaton $\tau$ = 1/2 $\pi$ f. We observed that the relaxation time as a function of absolute temperature (1000/T) gives a stright line which is well obayed by Arhenioud behavior. From the fittng parameters we have obtained activaton energy as E$_\alpha$ = 0.11 eV.

\subsection{Electric AC conductivity} 

To further understand the charge hoping and electrical properties we have investigated the AC conductivity in Sm$_2$CoMnO$_6$. The AC conductivity is calculated by using relation $\sigma_{ac} = \epsilon_0 \omega \epsilon^{\prime\prime}$.\cite{sing} Fig. 14a shows the variation of AC conductivity with frequency i.e. $\sigma_{ac}$ vs $f$ at some selective temperatures in the range 50 K to 300 K. It is evident from the figure that at low frequencies the conductivity is independent of frequency and gives a plateau region at all temperatures. In this region of frequency the conduction is mainly dominated by DC conductivity ($\sigma_{dc}$). However, at higher frequencies the conductivity increases with increasing frequency. It is further notable that the  plateau region marked by dc conductivity in the Fig. 14a extends to higher frequencies with increasing temperature. The frequency independent region also suggests that the hopping charge carriers are absent at low frequencies. The ac conductivity at high frequency in this case obey Jonscher's Universal Power Law given as follow:\cite{thakur}

\begin{eqnarray}
\sigma_{ac} = \sigma_{dc} + A\omega^n
\end{eqnarray}

when A is a temperature dependent constant, $\omega$ = 2$\pi$$f$ and $n$ is the power law exponent which generally varies between 0 and 1 depending upon temperature. The value of power law exponent $n$ represent the extent of interaction between mobile ions and lattice around. For non-interacting Debye system $n$ = 1 and with decreasing $n$ value the interaction is expected to increase between lattice surrounding and mobile ions. Further the constant A  gives the degree of polarizibility.

\begin{figure}[th]
	\centering
		\includegraphics[width=8cm]{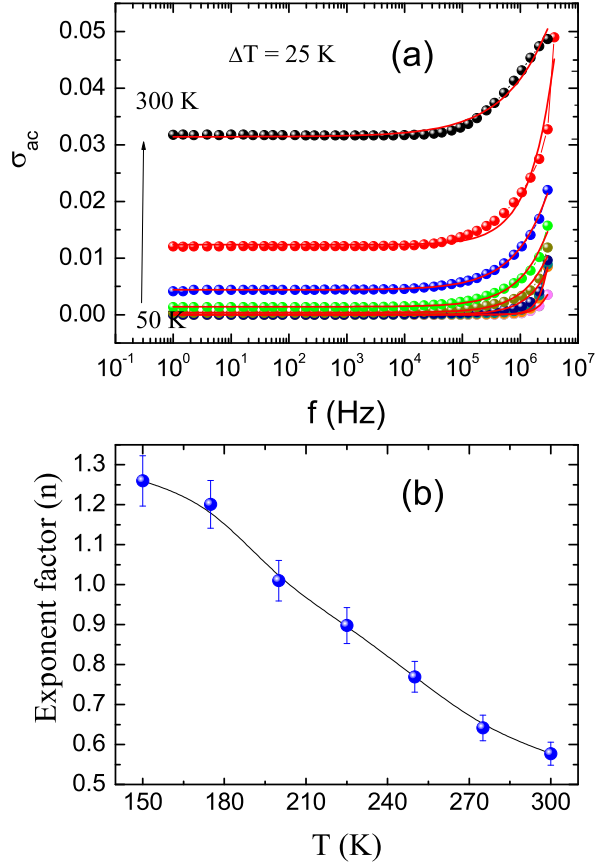}
\caption{(Color online) (a) Frequency dependence plot of the ac conductivity ($\sigma_{ac}$) for temperatures ranging from 50 K to 300 K are shown for Sm$_2$CoMnO$_6$. Solid red lines are fitting due to Eq. 11. (b) Shows the temperature variation of the power law exponent $n$.}
	\label{fig:Fig14}
\end{figure}

\begin{figure}[t]
	\centering
		\includegraphics[width=8cm]{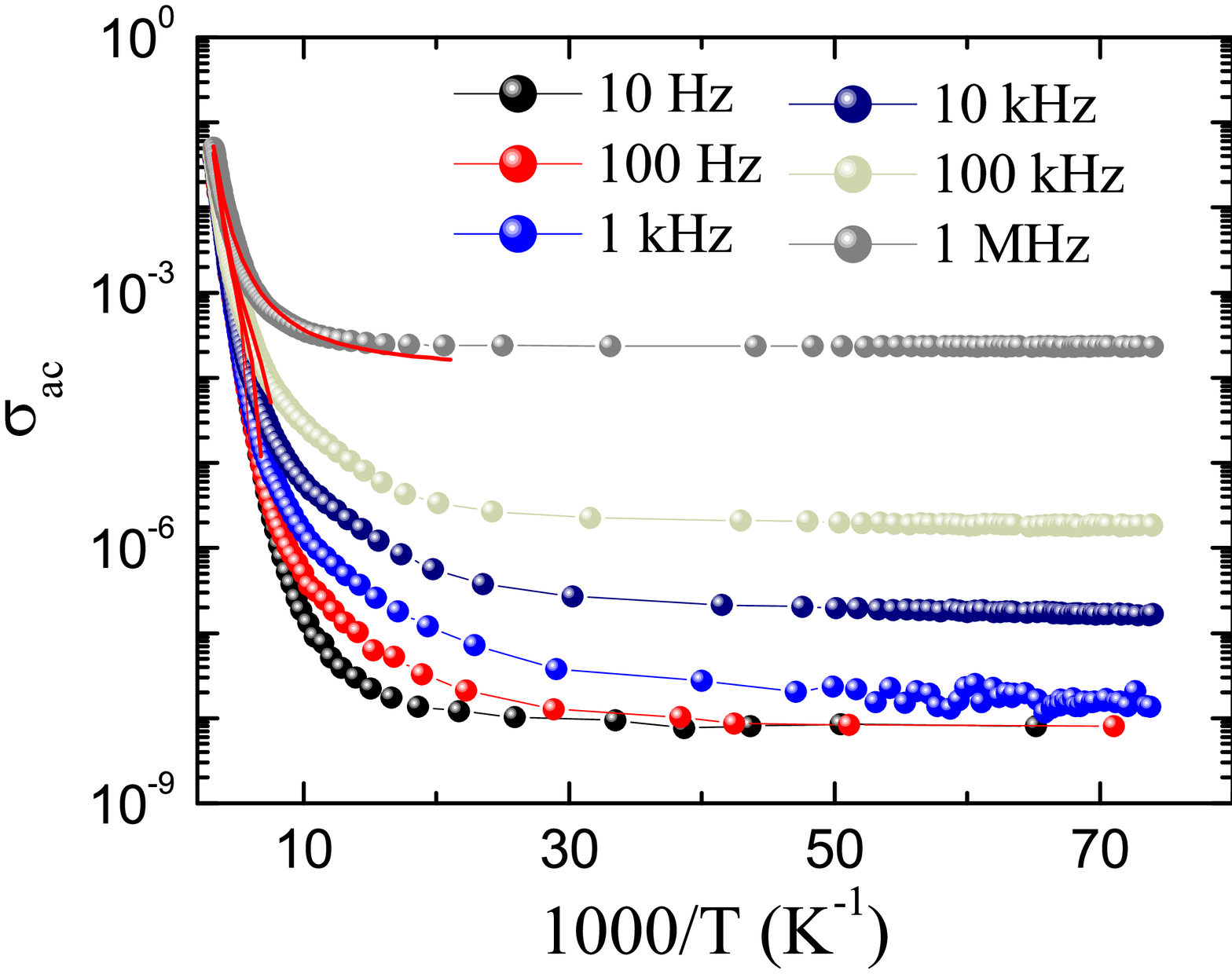}
\caption{(Color online) The variation of $\sigma_{ac}$ with absolute temperature (10$^3$/T) is shown for Sm$_2$CoMnO$_6$. The solid line are due to fitting with Eq. 12.}
	\label{fig:Fig15}
\end{figure}

The frequency dependent $\sigma_{ac}$  is fitted with the power law as shown in Fig. 14a. The conductivity data is fitted well  in full frequency range 1 Hz to 5.5 MHz. From the fitting parameters the exponent factor $n$ is obtained for all temperatures and is plotted in Fig. 14b. It ss evident from the figure that the $n$ value at room temperature is around 0.5 which increases with decreasing temperature. However we found that below 150 K the conductivity data does not obey the Power law behavior.

To further understand the conduction mechanism we have measured temperature dependent AC conductivity.  The variation of AC conductivity with absolute temperature i.e.  $\sigma_{ac}$ vs 10$^3$/T at some selective frequencies is shown in Fig. 15. We observed that with increasing frequency the conductivity increases.

\begin{eqnarray}
\sigma_{ac} = \sigma_{0}exp\left(\frac{-E_\alpha}{k_BT}\right)
\end{eqnarray}
where $\sigma_{0}$ is pre-exponent factor, $k_B$ is Boltzmann constant and $E_\alpha$ is the activation energy.

We have fitted the conductivity data with the Eq. 12. It is found that the conductivity data is well fitted at high temperatures for all frequency. At low temperature the AC conductivity is independent of temperature and form a steady region but does depend on frequency. From the fitting parameter we calculated the activation energy which decreases with increasing frequency  listed as for E$_{\alpha}$(1MHz) = 2.15 eV, E$_{\alpha}$(100 kHz) = 2.67 eV, E$_{\alpha}$(10 kHz) = 2.86  eV, E$_{\alpha}$(1 kHz) = 2.87 eV, E$_{\alpha}$(100 Hz) = 2.88 eV. 

\section{Conclusion}
Nano-crystalline Sm$_2$CoMnO$_6$ sample is successfully prepared by the sol-gel method. Structural, magnetic and dielectric properties are studies in detail. Structural study by x-ray diffraction confirms that the sample is in single-phase and chemically pure.  Rietveld refinement shows that Sm$_2$CoMnO$_6$ crystallized in monoclinic crystal structure and adopt \textit{P2$_1$/n} space group. Magnetization study reveals that  Sm$_2$CoMnO$_6$ is a ferromagnetic material that undergoes paramagnetic to ferromagnetic phase transition around 148 K marked as transition temperature ($T_c$). The effective magnetic moment obtained from experimental data is close to the theoretical value for the spin only system in the high spin state for Co/Mn sublattice. It is further observed that the Sm$^{3+}$ spin ordered in opposite direction to the Co/Mn sublattice thus giving rise to antiferromagnetic ordering at low temperature. Temperature dependent Raman study shows deviation from anharmonic behavior of frequency and line width for A$_{g1}$ Raman mode around $T_c$ which confirms the presence of spin-phonon coupling in Sm$_2$CoMnO$_6$. The dielectric response is studied in detail we observed that the material has a high dielectric constant at low temperature. The dielectric loss shows a relaxation process due to grains, which is thermally activated in nature with an activation energy of 0.14 eV. The impedance spectroscopy and electrical modulus study further confirm that the relaxation process is thermally activated in nature where the relaxation time follows Arrhenius behavior. AC conductivity has been studied with both frequency and temperature dependence. we found that the activation energy decreases with increasing frequency. Further, the exponent factor obtained from conductivity analysis shows that the system deviates from the Debye's model. 

\section{Acknowledgment}
We acknowledge MNIT Jaipur, India for XPS data and AIRF (JNU) for magnetic measurement facilities. We acknowledge UGC-DAE-Consortium Indore and Dr. V. G. Sathe for Raman data. We also acknowledge Dr. A. K. Pramanik for dielectric measurement and UPEA-II funding for LCR meter. We thank Saroj Jha
and Dr. Ruchita Pal for help in recording data. Author Ilyas Noor Bhatti acknowledge University Grants Commission, India for financial support.


\begin{thebibliography}{}

\bibitem{ndi} C. Gauvin-Ndiaye, A.-M. S. Tremblay, and R. Nourafkan, Phys. Rev. B \textbf{99}, 125110 (2019).
\bibitem{vilar} S. Yáñez-Vilar, E. D. Mun, V. S. Zapf, B. G. Ueland, J. S. Gardner, J. D. Thompson, J. Singleton, M. Sánchez-Andújar, J. Mira, N. Biskup, M. A. Señarís-Rodríguez and C. D. Batista, Phys. Rev. B \textbf{84}, 134427 (2011).
\bibitem{yang} K. Yang, D. I. Khomskii and H. Wu, Phys. Rev. B \textbf{98}, 085105 (2018).
\bibitem{dass} R. I. Dass and J. B. Goodenough, Phys. Rev. B \textbf{67}, 014401 (2003).
\bibitem{ndia} C. Gauvin-Ndiaye, T. E. Baker, P. Karan, É. Massé, M. Balli, N. Brahiti, M. A. Eskandari, P. Fournier, A.-M. S. Tremblay and R. Nourafkan, Phys. Rev. B \textbf{98}, 125132 (2018).
\bibitem{de} C. De and A. Sundaresan, Phys. Rev. B \textbf{97}, 214418 (2018).
\bibitem{ghara} S. Ghara, B.-G. Jeon, K. Yoo, K. H. Kim and A. Sundaresan, Phys. Rev. B \textbf{90}, 024413 (2014).
\bibitem{ilyas4} Ilyas Noor Bhatti, Rabindra Nath Mahato, Imtiaz Noor Bhatti and M.A.H. Ahsan, Materials Today: Proceedings, \textbf{17}, Part 1, 216 (2019).
\bibitem{fuh} H.R. Fuh and Y.P. Liu and Z.R. Xiao and Y.K. Wang, J. Magn. Magn. Mater. \textbf{357}, 7 (2004).
\bibitem{azu} M. Azuma, K. Takata, T. Saito, S. Ishiwata, Y. Shimakawa and M. Takano, J. Am. Chem. Soc. \textbf{127}(24), 8889 (2005).
\bibitem{choudhary} D. Choudhury, P. Mandal, R. Mathieu, A. Hazarika, S. Rajan, A. Sundaresan, U. V. Waghmare, R. Knut, O. Karis, P. Nordblad and D. D. Sarma, Phys. Rev. Lett. \textbf{108}, 127201 (2012).
\bibitem{rogado} N.S. Rogado, J. Li, A.W. Sleight and M.A. Subramanian, Advanced Materials \textbf{17}(18), 2225 (2005).
\bibitem{hoss2} A. Hossain, A. K. M. Atique Ullah, P. S. Guin and S. Roy, J Sol-Gel Sci Technol (2019), https://doi.org/10.1007/s10971-019-05054-8.
\bibitem{zhang} J. T. Zhang, X. M. Lu, X. Q. Yang, J. L. Wang, and J. S. Zhu, Phys. Rev. B \textbf{93}, 075140 (2016).
\bibitem{gau} C. Gauvin-Ndiaye, T. E. Baker, P. Karan, É. Massé, M. Balli, N. Brahiti, Phys. Rev. B \textbf{98}, 125132 (2018).
\bibitem{moon} J. Y. Moon, M. K. Kim, D. G. Oh, J. H. Kim, H. J. Shin, Y. J. Choi, and N. Lee, Phys. Rev. B \textbf{98}, 174424 (2018).
\bibitem{bisco1} J. Blasco, J. García, G. Subías, J. Stankiewicz, J. A. Rodríguez-Velamazán, C. Ritter, J. L. García-Muñoz and F. Fauth, Phys. Rev. B \textbf{93}, 214401 (2016).
\bibitem{bisco2}  J. Blasco, G. Subías, J. García, J. Stankiewicz, J. A. Rodríguez-Velamazán, C. Ritter and J. L. García-Muñoz, Solid State Phenomena \textbf{257}, 95 (2017).
\bibitem{bisco3} J. Blasco, J. L. García-Muñoz, J. García, G. Subías, J. Stankiewicz, J. A. Rodríguez-Velamazán and C. Ritter, Phys. Rev. B \textbf{96}, 024409 (2017).
\bibitem{ilyas1} Ilyas Noor Bhatti, Imtiaz Noor Bhatti, R. N. Mahato and M. A. H. Ahsan, Physics Letter A \textbf{383}, 2326 (2019).
\bibitem{ilyas2} Ilyas Noor Bhatti, R. N. Mahato, Imtiaz Noor Bhatti, and M. A. H. Ahsan, Physica B: Condensed Matter \textbf{558}, 59 (2019).
\bibitem{wliu} W. Liu, L. Shi, S. Zhou, J. Zhao, Y. Li and Y. Guo, J. Appl. Phys. \textbf{116}, 193901 (2014).
\bibitem{ilyas3} Ilyas Noor Bhatti, Imtiaz Noor Bhatti, R. N. Mahato and M. A. H. Ahsan, Ceramics International \textbf{46}, 46 (2020).
\bibitem{silva} R. X. Silva, M. C. Castro Junior, S. Yanez-Vilar M. S. Andujar, J. Mira, M. A. Senaris-Rodriguez and C. W. A. Paschoal, J. Alloy. Comp. \textbf{690}, 909 (2017).
\bibitem{nath} P. R. Mandal, R. C. Sahoo and T. K. Nath, Materials Research Express, \textbf{1}, 046108 (2014).
\bibitem{jwchen} J.-W. Chen, K. R. Chiou, A.-C. Hsueh and C.-R. Chang,  RSC Adv.\textbf{9}, 12319 , (2019).
\bibitem{jgjwang} G. J. Wang, C. C. Wanga, S. G. Huang, X. H. Sun, C. M. Lei, T. Li and L. N. Liu, AIP Advances \textbf{3}, 022109 (2013).


\bibitem{good} J. B. Goodenough, \textit{Magnetism and the Chemical Bond}, New York: Inter-Science (1976).
\bibitem{vasi} A. N. Vasiliev, O. S. Volkova, L. S. Lobanovskii, I. O. Troyanchuk, Z. Hu, L. H. Tjeng, D. I. Khomskii, H.-J. Lin, C. T. Chen, N. Tristan, F. Kretzschmar, R. Klingeler and B. Büchner, Phys. Rev. B \textbf{77}, 104442 (2008).
\bibitem{bhatti1} Imtiaz Noor Bhatti, R. S. Dhaka and A. K. Pramanik, Phys. Rev. B \textbf{96}, 144433 (2017).
\bibitem{bhatti2} Imtiaz Noor Bhatti, R. Rawat, A. Banerjee and A.K. Pramanik, J. Phys.: Condens. Matter \textbf{27}, 016005 (2014).
\bibitem{wang} X. Wang, W. Li, X. Wang, J. Zhang, L. Sun, C. Gao, J. Shang, Y. Hu and Q. Zhu, RSC Adv. \textbf{7}, 50753 (2017).
\bibitem{qiu} B. Qiu, W. Guo, Z. Liang, W. Xia, S. Gao, Q. Wang, X. Yu, R. Zhao and R. Zou, RSC Adv. \textbf{7}, 13340 (2017).
\bibitem{xia} H. Xia, D. Zhu, Z. Luo, Y. Yu, X. Shi, G. Yuan and J. Xie, Scientific Reports \textbf{3}, 2978 (2013).
\bibitem{ida} T. Hishida, K. Ohbayashi, and T. Saitoh, J. Appl. Phys. \textbf{113}, 043710 (2013).
\bibitem{sachoo} R. C. Sahoo, D. Paladhi and T. K. Nath, J. Magn. Magn. Mater. \textbf{436}, 77 (2017).
\bibitem{cao} Y. Cao, W. Li, K. Xu, Y. Zhang, T. Ji, R. Zou, J. Yang, Z. Qin and J. Hu, J. Mater. Chem. A \textbf{2}, 20723 (2014).
\bibitem{qliu} Q. Liu,  H. Yang,  H. Dong,  W. Zhang,  B. Bian,  Q. He,  J. Yang,  X. Meng,  Z. Tiana  and  G. Zhao, New J. Chem. \textbf{42}, 13096 (2018).
\bibitem{duan} D. Duan,  C. Hao,  W. Shi,  H. Wanga  and  Z. Sun, RSC Adv. \textbf{8}, 11289 (2018).
\bibitem{haung} Z.Huang, W. Zhou, C. Ouyang, J. Wu, F. Zhang, J. Huang, Y. Gao and J. Chu, Scientific Reports \textbf{5} 10899 (2015).
\bibitem{booth} R. J. Booth, R. Fillman, H. Whitaker, A. Nag, M. R. Tiwari, K. V. Ramanujachary,
J. Gopalakrishnan and S. E. Lofland, Mater. Res. Bull. \textbf{44}, 1559 (2009).
\bibitem{tzhang} T. Zhang, X.P. Wang, Q.F. Fang and X.G. Li, Appl. Phys. Rev. \textbf{1} 031302 (2014).
 \bibitem{rer} B. Raveau and Md. M. Seikh, \textit{Cobalt Oxides: From Crystal Chemistry to Physics},
John Wiley $\&$ Sons, 2012.
\bibitem{hoss1} A. Hossain, D. Ghosh, U. Dutta, P. S.Walke, N. E.Mordvinova, O. I. Lebedev, B. Sinha,  K. Pal, A. Gayen, A. K.Kundu, Md. M. Seikh, J. Magn. Magn. Mater. \textbf{444}, 68 (2017).
\bibitem{ilive} M. N. Iliev, M. V. Abrashev, A. P. Litvinchuk, V. G. Hadjiev, H. Guo and A. Gupta, Phys. Rev. B \textbf{75}, 104118 (2007).
\bibitem{meyer} C. Meyer, S. Hühn, M. Jungbauer, S. Merten, B. Damaschke, K. Samwer, and V. Moshnyaga, J. Raman Spectrosc. \textbf{48}, 46 (2017).
\bibitem{harish} Harish Kumar, V. G. Sathe and A. K. Pramanik, J. Magn. Magn. Mater. \textbf{478}, 148 (2019).
\bibitem{sandi} L. J. Sandilands, Y. Tian, K. W. Plumb, Y.-J. Kim, and K. S. Burch, Phys. Rev. Lett. \textbf{114}, 147201 (2015).
\bibitem{grana} E. Granado, A. García, J. A. Sanjurjo, C. Rettori, I. Torriani, F. Prado, R. D. Sánchez, A. Caneiro and S. B. Oseroff, Phys. Rev. B \textbf{60}, 11879 (1999).
\bibitem{lave} J. Laverdière, S. Jandl, A. A. Mukhin, V. Yu. Ivanov, V. G. Ivanov and M. N. Iliev, Phys. Rev. B \textbf{73}, 214301 (2006).
\bibitem{fang} L. Fang, H. Zhang, T. H. Huang, R. Z. Yuan and H. X. Liu, J. Mater. Sci. \textbf{40}, 533 (2005).
\bibitem{mansuri} Amantulla Mansuri, Ilyas Noor Bhatti, Imtiaz Noor Bhatti and Ashutosh Mishra, Journal of Advanced Dielectrics  \textbf{08}, No. 04, 1850024 (2018).
\bibitem{sing} A. Sing, A. Gupta and R. Chatterjee, Appl. Phys. Lett. \textbf{93} 022902 (2008).
\bibitem{thakur} S. Thakur, R. Rai, Igor Bdikin and M.A. Valente, Mater. Res. \textbf{19} 1 (2016).
\end{thebibliography}
\end{document}